\documentstyle[aps,prl,epsf,twocolumn,epsfig,floats]{revtex}

\begin{document}
\draft



\wideabs{

\title{Elecromagnetic generation and detection of dc/ac spin current} %
\author{Chia-Ren Hu\cite{email}}
\address{Department of Physics, Texas A\&M University,
College Station, TX 77843-4242}

\date{November 20, 2002}
\maketitle
\begin{abstract}
It is shown that in a heterostructure where screening is
eliminated, the Amp\'ere-Maxwell law of electrodynamics implies
that a dc or ac spinomotive force can be inudced with a time
rate of change of a transverse electric field, and the magnetic
analog of Amp\'ere law implies that a dc or ac spin current can
generate a transverse electric field at the same frequency
outside the spin-current channel. Both effects are quite weak
but may be of some usefulness.

\end{abstract}
} 

The term ``spintronics'' has appeared in sicentific literature
for almost a decade.~\cite{name} It is broadly defined to mean
any electronic application where the spin degree of freedom of
an electron is non-trivially utilized.~\cite{review} The word,
however, suggests another narrower definition, i.e., electronics
with the electron charge replaced by the electron spin. In this
narrower definition, an electric current is replaced by a spin
current, which means that the spin-up and -down electrons are
moving in opposite directions. (The direction of a spin current
is then that of the spin-up electrons. In a non-spin-polarized
conductor such a spin current is not accompanied by a net charge
current.) At the same time, an electromotive force ({\it emf}) is
replaced by a spinomotive force ({\it smf}), which pushes spin-up
and -down electrons in opposite directions with the same force
magnitude. (The direction of an {\it smf} is then that of the
forces acting on the spin-up electrons.) One may also define
such terms as spinoresistance, and, for an ac spin current,
spinocapacitance and spinoinductance, and even entertain such
notions as spinotransformer, spinorectifier, spinotransistor,
etc. Will this whole line of thoughts~\cite{note1} be as
practically useful as electronics? An important prerequesite for
an yes answer is to have a convenient power source for spin
current, i.e., a device to generate an {\it smf}. Recently, the
idea of a ``spin-cell'' has been introduced,~\cite{spincell}
which has to some extend achieved such a goal. However, in
electronics, by far the most convenient way to provide an {\it
emf} is through induction --- i.e., via Faraday's law,
$\nabla\times {\bf E} = -\,\partial {\bf B}/\partial t$.
It has the additional advantage of allowing either a dc or an
ac {\it emf} to be generated. One purpose of this letter is
to show that there is indeed an inductive way to generate an
{\it smf}, but not by Faraday's law, but rather by its magnetic
analog:
\begin{equation} %
\nabla\times {\bf H} = \partial {\bf D}/\partial t\;
(+ \,{\bf J}_e)\,,
\label{AmpereMaxwell} %
\end{equation} %
or the Amp\'ere-Maxwell law (with the current term enclosed in
brackets not important here).~\cite{note2} Unfortunately, as
we shall see, this process can only generate a very weak {\it
smf}, although probably sufficient for some purposes. A second
purpose of this letter is to show how the magnetic analog of the
Biot-Savart law, or equivalently, that of Amp\'ere's law,
\begin{equation} %
\nabla\times {\bf E} = -{\bf J}_m\,,
\label{magAmpere} %
\end{equation} %
allows a simple electromagnetic detection of a (dc or ac) spin
current, but the sensitivity is quite low, so only very large
spin currents can be detected this way. Still, we think that
this second idea can also have some usefulness, since observing
this effect would be a solid confirmation of the existence of a
spin current.

If electrons had only spin property but no charge property
(like neutrons), the inductive way to generate an {\it smf}
and the electromagnetic way to detect a spin current would
be already contained in a paper published by the author a
long time ago.~\cite{hu82} (That work was presented in the
context of superfluid $^3$He-$A_1$, so it can hardly reach the
spintronics community. The magnitudes examined there were even
smaller, due to the involvement of $^3$He nuclear magnetic
moment, rather than the electronic magnetic moment involved here,
and the very small superfluid fraction $\rho_s/\rho$, because
the $A_1$ phase of superfluid $^3$He exists in a narrow
temperature range near $T_c$ only.) The fact that electrons also
have charge property requires a modification of the idea
presented there. More specifically, it is the necessity to solve
the problem that an external electric field applied to a
conductor is screened, and cannot reach most of the electrons in
a conductor. This problem may be simply solved by fabricating a
heterostructure, which is made of alternating layers of
conductors and insulators, as is schematically illustrated in
Fig. 1.
\begin{figure}[htb] %
\centerline{\epsfig{figure=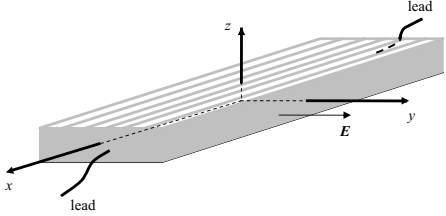,width=4.0in}}
\vspace{.2cm} %
\caption{A heterostructure made of alternating conducting
(grey) and insulating (white) layers (perpendicular to the $y$
axis) for generation and detection of spin current. The two end
faces perpendicular to the $x$ axis are coated with conducting
material so that leads attached to them are connected to all
conducting layers of the heterostructure.}
\end{figure} %
The thicknesses of these layers do not have to be strictly
uniform, and the interfaces need only be reasonably flat (just
to reduce surface scattering). Each conducting layer needs only
to be well connected and thinner than the screening length ---
hence it may be necessary to use conductors with low carrier
concentrations so that the screening length can encompass
several atomic layers. Each insulating layer also needs to be so
thick as to prohibit tunneling between the neighboring conducting
layers, but not too thick in order to allow many conducting
layers to be packed into a convenient size. In this way
screening can be essentially suppressed for electric field
applied perpendicular to the layers. The total number of
conducting layers must be such that their total thickness is
macroscopic, so that the device can drive a macroscopic spin
current. Each conducting layer can be taken as a $L_x\times L_z$
rectanglular sheet, as shown in Fig. 1, where the edges of these
sheets are parallel to the $x$ ans $z$ axes. The $x = \pm
(1/2)L_x$ end faces of this heterostructure must still be coated
with a thin layer of conducting material --- the two $L_y\times
L_z$ rectangular conducting sheets at $x = \pm (1/2)L_x$ in Fig.
1, so that one lead connected to one end face can send spin
current into all conducting layers in the heterostructure, and
another lead connected to the other end face can take spin
current out of them. The leads then serve as the two terminals
of this ``spinovoltage generator''.~\cite{note3}

To generate a dc {\it smf} ${\cal E}_m$ across these two
terminals, one must subject this heterostructure to a
uniform constant time rate of change of electric field $dE_y/dt$
in the $y$ direction, i.e., perpendicular to the conducting
sheets in the heterostructure. Let the spin-up(down) electron
density in the conducting sheets be denoted as $n_{\uparrow}$
($n_{\downarrow}$) (defined with respect to the $z$ axis). Then
the spin-up electrons contribute a magnetization $M_{z\uparrow}
= -\mu_B n_{\uparrow}$ in the conducting sheets (in the $z$
direction), and the spin-down electrons contribute a
magnetization $M_{z\downarrow} = +\mu_B n_{\downarrow}$ in the
conducting sheets, where $\mu_B$ is the Bohr magneton. Assuming
that the electron densities are uniform in the conducting
sheets, $M_{z\uparrow}$ ($M_{z\downarrow}$) is equivalent to a
surface magnetic-charge density $\sigma_{m\uparrow} =
M_{z\uparrow}$ ($\sigma_{m\downarrow} = M_{z\downarrow}$) at $z
= +L_z/2$, and another surface magnetic-charge density
$-\sigma_{m\uparrow} = -M_{z\uparrow}$ ($-\sigma_{m\downarrow} =
-M_{z\downarrow}$) at $z = -L_z/2$. If the system is unpolarized,
these magnetic charge densities exactly cancel each other. But
that does not prevent opposite magnetic charges from moving in
opposite directions if they are subject to a magnetic field.
This magnetic field is induced by the applied $dE_y/dt$
according to the Amp\'ere-Maxwell law (Eq.~\ref{AmpereMaxwell})
and is $H_{x\pm} = \pm \epsilon_0\,(L_z/2)\,(dE_y/dt)$ at $z =
\pm L_z/2$ where we have neglected the dielectric property of
the conducting sheets. The forces acting on all magnetic charges
at $z = \pm L_z/2$ due to spin-up (or down) electrons are then in
the same direction, and have the magnitude $F_{\uparrow
(\downarrow)} = \pm\sigma_{m\uparrow (\downarrow)} L_x L_y
H_{x\pm}$ This leads to a volume force acting on all spin-up
(down) electrons uniformly in $-x$ ($+x$) directions (if
$dE_y/dt>0$), and is therefore an {\it smf}. We propose to
artificially measure it in terms of the familiar unit, volt (V)
(which actually means to measure the potential energy gain per
electron in $e$V). Then the {\it smf} induced in this
device (in the $x$ direction) is %
\begin{equation} %
{\cal E}_m = - \epsilon_0\,(\mu_B/e)\,L_x\,(dE_y/dt)\,. %
\label{dcsmf}
\end{equation} %

One can also generate an ac {\it smf} this way, if one
replaces the constant $dE_y/dt$ by the time rate of
change of an ac electric field in the $y$ direction: $E_y(t) =
E_0\cos\omega t$. Then one has
\begin{equation} %
{\cal E}_m(t) = \epsilon_0\,(\mu_B/e)\,L_x\,E_0\,\omega\,\sin(\omega t)\,. %
\label{acsmf}
\end{equation} %

To estimate the magnitude of this ac {\it smf}, one needs
to use $\mu_B = 9.27410\times 10^{-24}{\rm J/T}$. This $\mu_B$
must still be multiplied by the magnetic permeability of vacuum,
$\mu_0$, before it should be used in Eq.~\ref{acsmf}, since a
magnetic dipole moment ${\bf \mu}$ appears in such formulas as
(energy) $U = -{\bf \mu}\cdot{\bf B}$, and (torque) ${\bf N} =
{\bf\mu}\times{\bf B}$, which contains ${\bf B}$ rather
than ${\bf H}$, so it does not yet have the unit of magnetic
charge times length until it is multiplied by $\mu_0$, as a
magnetic charge $q_m$ should appear in the formula (force) ${\bf
F} = q_m {\bf H}$, with ${\bf H}$ appearing rather than
${\bf B}$.~\cite{Jackson}
Knowing that the effect is very weak, we insert in our estimate
rather large but presumably still achievable values of $L_x$,
$E_0$ and $\omega$  viz., $10$ cm, $10^6$ V/m, and $10^{12}$ Hz,
just to see what is the largest possible magnitude we can obtain.
We then obtain the magnitude of the induced ac {\it smf} to be
about $6.44\times 10^{-5}$ V. This is still very small, but
probably already sufficient for some purposes. One could still
connect many such devices in series in order to get a larger
{\it smf} (by another factor of 10 or even 100). Thus one could
conceivably generate a several-milli-volt {\it smf} through such
an inductive process. Since the frequency used already
corresponds to a wave length of about 2 mm, we need to limit
$L_z$ to be below about 1 mm, so that the electric field
experienced by all electrons in the conducting sheets can be in
phase. The length $L_y$ has no limitations except that it should
not be too big, so it can be, say, 10 cm. The heterostructure
should then be in a microwave cavity designed so that the
(standing) wave vector is in the $z$ direction. (The wave vector
should not be in the $x$ direction since the so-generated {\it
smf} is proportional to $L_x$, so we need $L_x$ to be as large
as possible.) One could in principle also rotate the
heterostruction in a constant electric field to get the
electrons to see an ac field, but one can hardly rotate the
device at $\omega = 10^{12}$ Hz, so this alternative approach is
impractical.

The design shown in Fig.~1, with slight modification, can also
serve as a detector of a dc or ac spin current, but we shall see
that the sensitivity is quite low, so only very large spin
currents can be detected this way. One needs only remove the
applied electric field to the device, and send the spin current
through the heterostructure via the two terminals, and then
detect an electric field in the $y$ direction that is generated
by the spin current outside the spin-current channel.
The relevant law involved is now the
magnetic analog of the Amp\'ere law, Eq.~\ref{magAmpere}. As we
have already explained, the magnetization due to all spin-up (or
-down) electrons in a conducting sheet is equivalent to a
distribution of magnetic charges of opposite signs on the two
end faces at $L_z = \pm L_z/2$. Thus the motion of these
electrons implies opposite magnetic charge currents at $z = \pm
L_z/2$. Together, they may be viewed as a ring magnetic charge
current in the $xz$ plane which can generate an electric field
in the $y$ direction. Looking at the conducting sheets from $y =
+\infty$ the ring magnetic charge current on each conducting
sheet is clockwise from both spin-up and -down electrons (since
they are moving in opposite directions), for a spin current in
the $+x$ direction. Thus they both generate electric field in
the $+y$ direction. (Notice the minus sign on the right hand
side of Eq.~\ref{magAmpere}.) Since each conducting sheet is
very thin, the ring magnetic charge current in it may be treated
as a line current. Then It is easy to work out the magnetude of
the electric field generated along the $y$ axis:
\begin{equation} %
E_y(0,0,y) = \sum_{n = - n_0}^{n_0}\frac{\mu_B\,
 (n_{\uparrow}v_{\uparrow} -  n_{\downarrow}v_{\downarrow})\,
  t\,L_z/2}  {2\,\pi\,[(y-nd)^2
 + L_z^2/4]}\,,
\label{Eyonyaxis} %
\end{equation} %
where $t$ and $d$ (asummed $>> t$) are the thicknesses of each
conducting and insulating sheet, respectively, and
($v_{\uparrow}$, $v_{\downarrow}$) denote the velocities of the
spin-(up,down) electrons, respectively. We have also assumed that
there are $2n_0 + 1$ conducting sheets symmetrically placed
between $y = -n_0 d$ and $y = +n_0 d$. Since $d$ is still very
small, we can approximate the above sum by an integral. It gives
\begin{eqnarray} %
E_y&&(0,0,y) = \frac{\mu_B\,t\,(n_{\uparrow}v_{\uparrow} -
n_{\downarrow}v_{\downarrow})}{\pi\,d}\times\nonumber\\
&&\left[\arctan\left(
\frac{y + n_0 d}{L_z/2}\right) - \arctan\left(\frac{y -
n_) d}{L_z/2}\right)\right]\,.
\label{Ey} %
\end{eqnarray} %
For $e(n_{\uparrow}v_{\uparrow} - n_{\downarrow}v_{\downarrow})
= 10^7$ A/m$^2$, and $d = 10 t$, we calculate the front
factor to be $7.274\times 10^{-5}$ V/m. This is a weak but
observable electric field, but the very large spin current used
in this calculation can hardly be generated by the induction
method discussed above. However, in a strongly polarized or half
metallic ferromagnetic metal, a very large polarized current can
be generated with an ordinary {\it emf}. If such a current is
sent through the present device, a detectible electric field
should be generated as shown here, besides the magnetic field
which must also occur because a spin-polarized current is still
a charge current. Observing this electric field would then be a
definitive way to confirm the existence of a spin-polarized
current.

Integrating the above electric field along the whole
$y$ axis, we find the total voltage drop along this axis to be:
\begin{equation} %
\Delta V \equiv\int_{-\infty}^{\infty}E_y dy = \mu_B\,
(n_{\uparrow}v_{\uparrow} - n_{\downarrow}v_{\downarrow})\,
(t/d)\,L_y\,.
\label{DeltaV} %
\end{equation} %
For the same spin current and $t/d$ considered in the previous
paragraph, and let $L_y = 10$ cm, we obtain $\Delta V =
7.274\times 10^{-6}$ V, which is a detectible voltage difference,
but it is cetainly a very weak one. If the spin current is ac,
then the induced electric field is also ac at the same
frequency. One can then possibly use a resonance technique to
detect this electric field with an improved sensitivity.

We conclude this work with some remarks: (1) The
conducting sheets can not be made of singlet superconductors,
since if the spin-up electrons and spin-down electrons form
bound Cooper pairs then they can not flow in opposite
directions. p-wave superconductors do not have this problem,
and it has the advantage of contributing no internal
spinoresistance, so it has the potential of generating
larger spin current, but one needs to worry about suppressed
order parameter in sheets much thinner than their coherence
length. So low-carrier-density normal conductors with low
resistance and low magnetic and spin-orbit scattering centers
are still the best bet for making this device. (2)
spinoresistance can result from both scattering (which reduces
electrons' forward momentum) and spin conversion. So in most
good conductors it should be of the same order of magnitude as
its usual resistance unless the conductor has a lot of magnetic
and/or spin-orbit scattering centers, in which case its
spinoresistance should be larger than its usual resistance. (3)
If the spin-up and -down currents can be separated in a short
section of the spin circuit using half-metallic conductors, and
the usual capacitors, inductors, transformers, rectifiers, or
transistors, etc., are inserted in the two separate circuit
branches, one would have created a spinocapacitor, a
spinoinductor, a spinotransformer, a spinorectifier, or a
spinotransistor, etc., but it doesn't appear that they can allow
spin circuits to maintain their advantages over the usual charge
circuits. Thus more clever ideas are needed before spin circuits
can compete with the usual charge circuits for usefulness.

The author wishes to thank his colleague, Joe Ross, for useful
discussions. This work was supported by the Texas Center for
Superconductivity and Advanced Materials at the University of
Houston.


\begin{references}
\bibitem{note1} We suggest calling this whole line of thoughts
``orthospintronics'', and referring to all other spintronic
ideas and applications as ``paraspintronics''.
\bibitem{note3} This heterostructure may be also fabricated as a
semiconductor multi-quantum-well structure. However, in order
for the device to work at room temperatures, a very wide gap
semiconductor must be used as the material between the quantum
wells. The quantum wells must also be doped to serve as the
conducting sheets. Their electron density can be controlled to
ensure that the width of the quantum wells are less than the
screening length.
\end{references}
\end{document}